\documentclass[12pt]{article}
\usepackage{amssymb,diagrams}
\newcommand{\be}{\begin{eqnarray}}
\newcommand{\ee}{\end{eqnarray}}
\newcommand{\A}{{\cal A}}
\renewcommand{\O}{\Omega}
\newcommand{\pa}{\partial}
\renewcommand{\d}{{\rm d}}
\newcommand{\D}{{\rm D}}
\newcommand{\Dc}{{\cal D}}
\newcommand{\dl}{{\delta}}
\title{\bf Bicomplexes and Integrable Models}
\date{  }
\author{A. Dimakis \\ Department of Mathematics, University of the Aegean \\
        GR-83200 Karlovasi, Samos, Greece \\ dimakis@aegean.gr
        \\[2ex]
        F. M\"uller-Hoissen \\ Max-Planck-Institut f\"ur Str\"omungsforschung \\
        Bunsenstrasse 10, D-37073 G\"ottingen, Germany \\
        fmuelle@gwdg.de }

\begin{document}
\renewcommand{\theequation} {\arabic{section}.\arabic{equation}}

\maketitle

\begin{abstract}
We associate bicomplexes with several integrable models in 
such a way that conserved currents are obtained 
by a simple iterative construction. Gauge transformations and dressings 
are discussed in this framework and several examples are presented, 
including the nonlinear Schr\"odinger and sine-Gordon equations, and some 
discrete models.
\end{abstract}

\section{Introduction}
\setcounter{equation}{0}
Let $D = d + A$ be the covariant exterior derivative associated
with a connection 1-form $A$.
The integrability condition of the linear equation $D \chi = 0$
for a vector valued function $\chi$ is the zero curvature condition 
$F = d A + A \wedge A = 0$ since $D^2 \chi = F \, \chi$. In two 
dimensions where $A = - U \, d x - V \, d t$ with matrices $U$ and $V$ 
depending on coordinates $x,t$, the zero curvature condition takes the form
$U_t - V_x + [U , V] = 0$ (cf \cite{Fadd+Takh87}, for example)
which can be rewritten in the form of a Lax equation.
Soliton equations and integrable models are known to possess such a
zero curvature formulation with a connection (i.e., $U$ and $V$)
depending on a parameter, say $\lambda$ (cf \cite{Fadd+Takh87}, for example).
\vskip.2cm

This geometric formulation of integrable models is easily extended
to generalized geometries, in particular in the sense of noncommutative
geometry where, on a basic level, the algebra of differential forms
on a manifold is generalized to a differential calculus over an 
associative algebra $\cal A$ (for which the algebra of smooth functions
on a manifold is an example) \cite{DMH96}.
\vskip.2cm

Recent work \cite{DMH00a,DMH99b,DMH99c} shows that for many integrable 
models there is a zero curvature formulation in which 
the linear system appears naturally in a form which depends {\em linearly} on 
the spectral parameter $\lambda$. However, translating the linear system into 
the form $\pa \chi/\pa x = U(x,t,\lambda) \, \chi$ and 
$\pa \chi/\pa t = V(x,t,\lambda) \, \chi$, as considered in \cite{Fadd+Takh87},
for example, one usually ends up with a {\em nonlinear} dependence of $U$ 
and $V$ on $\lambda$. An example in \cite{Fadd+Takh87} for which
$U$ and $V$ are linear in $\lambda$ is the N-wave model (see p. 309). A more
important example where the connection depends linearly on $\lambda$ is 
provided by the self-dual Yang-Mills equations from which
many integrable models can be obtained by a reduction procedure \cite{sdYM_red}. 
In such cases we have $D = \delta - \lambda \, \d$ 
where $\delta$ and $\d$ are anticommuting linear 
maps satisfying $\d^2 = 0 = \delta^2$. They constitute what is known 
as a {\em bicomplex} (and need not satisfy the Leibniz rule as in the case of 
a bidifferential calculus).
The linear system then reads
\be
     \delta \chi = \lambda \, \d \chi \; .
\ee
The most interesting point concerning this equation is not its simplicity 
in the dependence on $\lambda$, but rather the fact that expressing
$\chi$ as a power series in $\lambda$ leads in a very simple 
way to the conserved densities of the respective model. Moreover, behind this is
a general iterative construction of $\delta$-closed elements of a bicomplex. 
This is explained in more detail in section 2 which somewhat generalizes
the framework of our previous papers. In particular, a modification of the above 
linear system by adding an inhomogeneous term is necessary, in general.
\vskip.2cm

Applied to chiral models, the iterative construction of ``generalized conserved 
densities" in the sense of $\delta$-closed elements of a bicomplex is precisely
the construction of non-local conserved charges due to Br{\'e}zin, Itzykson,
Zinn-Justin and Zuber \cite{BIZZ79}. Our previous and the present work shows 
that the same method applies to most of the known soliton equations and 
integrable models (and perhaps to all of them). Surprisingly, in several cases
the apparently {\em nonlocal} construction leads to {\em local} conserved currents 
and charges, rather than nonlocal ones. 
\vskip.2cm

In section 3 we apply gauge transformations and dressings to some (trivial) 
bicomplexes. Since we consider two separate ``generalized covariant derivatives" 
instead of one depending on a (spectral) parameter, a gauge transformation 
can be applied to just one 
of them. Such a dressing transformation deforms the bicomplex in a relatively 
simple way and the bicomplex conditions lead to equations for the transformation 
map. In this way one recovers several integrable models. Applying a gauge 
transformation simultaneously to both maps, $\d$ and $\delta$, results in an equivalence 
transformation of the bicomplex, of course. We present several examples, including 
the nonlinear Schr\"odinger equation, the sine-Gordon equation and its discrete 
version, as well as a Toda field theory and a corresponding discretization. 
 Furthermore, we briefly discuss generalizations of the self-dual Yang-Mills 
equations and reductions in the bicomplex framework.
Section 4 contains some conclusions.

\section{Weak bicomplexes and associated linear equation}
\setcounter{equation}{0}
Let $ M = \bigoplus_{r \geq 0} M^r $
be an ${\mathbb{N}}_0$-graded linear space (over ${\mathbb{R}}$ or 
${\mathbb{C}}$) and $\d , \delta \, : \, M^r \rightarrow M^{r+1}$, 
$\rho \, : \, M^r \rightarrow M^r$ linear maps satisfying
\be
   \rho \, \d^2 = 0 \, , \qquad  \delta^2 = 0 \, , \qquad  
   \delta \, \d + \rho \, \d \, \delta = 0 \; .   \label{bicomplex_cond}
\ee 
Then $(M,\d,\delta, \rho)$ is called a {\em weak bicomplex}. 
If $\rho$ is the identity map, then $(M,\d,\delta)$ is a {\em bicomplex}.
In terms of $\d_\lambda = \delta - \lambda \, \d$ with a constant $\lambda$,
the three bicomplex equations can then be combined into the single 
condition $\d_\lambda^2 = 0$ (for all $\lambda$).
\vskip.2cm

We are interested in the case where the weak bicomplex maps depend on certain objects 
(e.g., functions or operators) in such a way that the above bicomplex equations are 
satisfied if these objects are solutions of some (e.g., differential or operator) 
equations. Of particular interest are those cases where the bicomplex conditions become
equivalent to a certain (e.g., nonlinear partial differential) equation.
\vskip.2cm

We assume that, for some $s \in {\mathbb{N}}$, there 
is a (nonvanishing) $\chi^{(0)} \in M^{s-1}$ with 
\be
   \rho \, \d J^{(0)} =0 \quad \mbox{where} \quad 
      J^{(0)} = \delta \chi^{(0)} \; .
\ee
Let us define 
\be 
      J^{(1)} = \d \chi^{(0)} \; . 
\ee
Then 
\be
     \delta J^{(1)} = - \rho \, \d \delta \chi^{(0)} = 0 \; .
\ee
If the $\delta$-closed element $J^{(1)}$ is $\delta$-exact, then 
\be
    J^{(1)} = \delta \chi^{(1)}
\ee 
with some $\chi^{(1)} \in M^{s-1}$. Next we define
\be
   J^{(2)} = \d \chi^{(1)} \; . 
\ee
Then 
\be
   \delta J^{(2)} = - \rho \, \d \delta \chi^{(1)} = - \rho \, \d J^{(1)} 
   = - \rho \, \d^2 \chi^{(0)} = 0 \; .
\ee
If the $\delta$-closed element $J^{(2)}$ is $\delta$-exact, then 
\be
    J^{(2)} = \delta \chi^{(2)}
\ee
with some $\chi^{(2)} \in M^{s-1}$. 
This can be iterated further and leads to a (possibly infinite) 
chain (see Fig.~1) of elements $J^{(m)}$ of $M^s$ and 
$\chi^{(m)} \in M^{s-1}$ satisfying
\be
         J^{(m+1)} = \d \chi^{(m)} = \delta \chi^{(m+1)}   \; .
\ee
More precisely, the above iteration continues from the $m$th to the
$(m+1)$th level as long as $\delta J^{(m)} = 0$ 
implies $J^{(m)} = \delta \chi^{(m)}$ with an element 
$\chi^{(m)} \in M^{s-1}$. Of course, there is no obstruction to the 
iteration if $H^s_\delta (M)$ is trivial, 
i.e., when all $\delta$-closed elements of $M^s$ are $\delta$-exact.
In general, the latter condition is too strong, however, though
in several examples it can be easily verified \cite{DMH00a}.
\vskip.2cm

Introducing 
\be
   \chi = \sum_{m \geq 0} \lambda^m \, \chi^{(m)}
\ee
with a parameter $\lambda$, the essential ingredients of the above iteration 
procedure are summarized in
\be
    \delta (\chi - \chi^{(0)}) = \lambda \, \d \, \chi     \label{bc-linear}
\ee
which we call the {\em linear equation} associated with the 
bicomplex.\footnote{Independent of the choice of $\rho$, we are led to the same 
form of the linear equation (\ref{bc-linear}). The map $\rho$ enters via the 
initial condition $\rho \, \d \delta \chi^{(0)} =0$. The linear equation then 
implies $[ \delta^2 - \lambda \, (\delta \d + \rho \, \d \delta) 
+ \lambda^2 \rho \, \d^2] \chi = (\delta^2 - \lambda \, \rho \, \d \delta) \chi^{(0)}
=0$.}

\vskip.2cm

\small
\diagramstyle[PostScript=dvips]
\begin{diagram}[notextflow]
M^{s-1}&       &               &\chi^{(0)}&         &       &               &\chi^{(1)}&         &       &      \\
\dTo   &       &\ldTo^\dl      &          &\rdTo^\d &       &\ldTo^\dl      &          &\rdTo^\d &       &      \\
M^s    &J^{(0)}&               &          &         &J^{(1)}&               &          &         &J^{(2)}&\cdots\\
\dTo   &       &\rdTo^{\rho \d}&          &\ldTo^\dl&       &\rdTo^{\rho \d}&          &\ldTo^\dl&       &      \\
M^{s+1}&       &               & 0        &         &       &               & 0        &         &       &      
\end{diagram}
\normalsize

\begin{center}
{\bf Fig. 1}  \\
The iterative construction of $\delta$-closed elements $J^{(m)} \in M^s$. 
\end{center}

\vskip.2cm

Applying $\delta$ to the bicomplex linear equation and using (\ref{bicomplex_cond}), we find
\be
     \rho \, \d \delta \chi = 0  \; .
\ee
Hence $\chi$ has to satisfy the same condition which we put on the initial data $\chi^{(0)}$. 
We may thus think of the above iteration as a discrete process in the space of solutions of 
this equation.
\vskip.2cm

Now we can turn things around. Given a bicomplex, we may start with the linear 
equation (\ref{bc-linear}). Let us suppose that it admits a (non-trivial)
solution $\chi$ as a (formal) power series in the parameter $\lambda$:
\be
    \chi = \sum_{m = 0}^N \lambda^m \, \chi^{(m)}
\ee
with $N \in \mathbb{N} \cup \{ \infty \}$. The linear equation leads to 
\be
   \delta \chi^{(m)} = \d \chi^{(m-1)} \, , \quad m=1, \ldots, N \, , \qquad 
       \d \chi^{(N)} = 0 
\ee 
where the last equation has to be dropped if $N = \infty$. As a consequence,
the $J^{(m+1)} = \d \chi^{(m)}$ ($m=0, \ldots, N-1$) are 
$\delta$-exact. Even if the cohomology $H^s_\delta (M)$ is {\em not} trivial,
the solvability of the linear equation ensures that the $\delta$-closed $J^{(m)}$
appearing in the iteration are $\delta$-exact.\footnote{If the cohomology 
condition holds, then the above iterative procedure provides us with a 
solution of the linear equation. Otherwise we have to show that the linear 
equation has a sufficiently nontrivial solution.}
This observation somewhat generalizes the framework of our previous papers 
and indeed appears to be necessary in order to cover examples like the 
nonlinear Schr\"odinger equation (see the following section).
\vskip.2cm

A priori, the mathematics presented above has little to do with conservation laws. 
However, formulating integrable models like KdV, KP, chiral models 
and the like in the bicomplex framework demonstrated that the $\delta$-exact
$J^{(m)}$ (where $s=1$) are directly or somewhat indirectly related to the known conserved 
densities of the respective models \cite{DMH00a,DMH99b,DMH99c}.
This is also confirmed by the examples treated in the following section. 
\vskip.2cm

The features usually attributed to {\em soliton equations} demand a high level of order 
and predictability, in complete contrast with chaotic systems which, in this sense, 
form the dark side of nonlinear dynamics. Soliton systems were found to possess an 
infinite set of conservation laws. This was taken as a (partial) explanation for 
the high order of simplicity of their scattering behaviour. If there is an infinite
chain of independent $\delta$-exact elements in a bicomplex associated with some
(integro-differential, difference, operator) equation, this is certainly also
a property expressing a high degree of order. In this sense the above structure 
should also be of interest beyond the context of integrable models.
\vskip.2cm

The freedom which enters the formalism through the possible choice of a map $\rho$ 
different from the identity and also the possibility of considering $s > 1$ will not 
be explored in this work. Hence, in the following we restrict our considerations to 
the simpler structure of a bicomplex and the case where $s=1$.
\vskip.2cm

In the examples which we present in the following sections, the bicomplex space is always
chosen as $M = M^0 \otimes \Lambda_n$ where $\Lambda_n = \bigoplus_{r=0}^n \Lambda^r$
is the exterior algebra of a (complex) $n$-dimensional vector space with a basis $\xi^r$,
$r=1, \ldots, n$, of $\Lambda^1$. It is then sufficient
to define the bicomplex maps $\d$ and $\delta$ on $M^0$ since via 
\be
 \d (\sum_{i_1,\ldots,i_r =1}^n \phi_{i_1 \ldots i_r} \, \xi^{i_1} \cdots \xi^{i_r} ) 
 = \sum_{i_1,\ldots,i_r =1}^n ( \d \phi_{i_1 \ldots i_r} ) \, \xi^{i_1} \cdots \xi^{i_r}
\ee
(and correspondingly for $\delta$) they extend as linear maps to the whole of $M$.
In the case of $\Lambda_2$ we denote the two basis elements of $\Lambda^1$ as $\tau, \xi$.

\section{Gauge transformations and dressings of bicomplexes}
\setcounter{equation}{0}
Let $(M,\d,\delta)$ be a bicomplex. 
A {\em gauge transformation} is a map of this bicomplex 
to another bicomplex $(M,\d',\delta')$ induced by an isomorphism
$g$ of $M$ such that
\be
    \d' \phi = g^{-1} \d (g \phi) \, , \qquad
    \delta' \phi = g^{-1} \delta (g \phi) \; .
\ee
Indeed, it is easily verified that $\d'$ and $\delta'$ satisfy 
the bicomplex conditions (\ref{bicomplex_cond}) (with $\rho = id$) 
if $\d$ and $\delta$ satisfy them.
\vskip.2cm

There are at least two simple ways to deform a bicomplex such that 
two of the bicomplex conditions (\ref{bicomplex_cond}) remain 
satisfied. In both cases we leave one of the two
bicomplex maps, say $\delta$, unchanged.
We call transformations of this kind {\em dressing transformations}.
\vskip.2cm

The first way is to transform $\d$ to
\be
  \tilde{\D} \phi = \d \phi +[\delta, v] \phi 
                  = \d \phi +\delta (v \phi)- v \, \delta \phi
                                     \label{dr2}
\ee
where $v$ is a linear map $M \rightarrow M$.
Then
\be
    \tilde{\D} \delta + \delta \tilde{\D} = \d \delta + \delta \d = 0
\ee
using $\delta^2 = 0$, so that all bicomplex equations besides 
$\tilde{\D}^2 =0$ are identically satisfied. The remaining condition 
takes the form
\be
  \d \delta (v \phi) - \d (v \delta \phi) - \delta (v \delta \phi) 
  + \delta (v \d \phi) - v \delta \d \phi + v \delta (v \delta \phi) = 0
      \; .
\ee
The problem is now to find $\d$ and $\delta$ such that the last equation
reduces to an interesting equation for $v$ independent of $\phi$.
\vskip.2cm
\noindent
{\em Example: The KP equation.} Let 
$M = C^\infty(\mathbb{R}^3) \otimes \Lambda_2$.
In terms of coordinates $t,x,y$ on $\mathbb{R}^3$ we define bicomplex 
maps $\d$ and $\delta$ via
\be
 \d f = (f_t-f_{xxx}) \, \tau + {1\over2}(f_y-f_{xx}) \, \xi  \, , \quad
 \delta f = {3\over 2}(f_y + f_{xx}) \, \tau + f_x \, \xi
\ee
for $f \in C^\infty(\mathbb{R}^3) = M^0$. The bicomplex equations 
(\ref{bicomplex_cond}) are then identically satisfied.
Deforming $\d$ to
\be
   \D' f &=& \d f + \delta (v f)- v \delta f  \nonumber \\
         &=& [f_t-f_{xxx}+{3\over 2}(v_y+v_{xx})f+3 v_x f_x] \, \tau 
		     +{1\over2}(f_y-f_{xx}+ 2 v_x f) \, \xi  
\ee
with $v \in M^0$ (which, by multiplication, acts linearly on $M$), 
$\D'{}^2=0$ becomes
\be
  v_{xt} - {1 \over 4} v_{xxxx} + 3 v_x v_{xx} - {3 \over 4} v_{yy} = 0
\ee
which is equivalent to the KP equation for the field $u=-v_x$.
           {  }   \hfill   \rule{5pt}{5pt}
\vskip.2cm

The second kind of dressing transformation is to transform $\d$ to
\be
    \D \phi = G^{-1} \d (G \phi)             \label{dr1}
\ee
where $G$ is an isomorphism of $M$. Then 
$\D^2 \phi = G^{-1} \d^2 (G \phi) = 0$ so that all bicomplex equations 
besides $\delta \D + \D \delta = 0$ are identically satisfied. 
The remaining condition results in the following equation involving $G$,
\be
    \delta [ G^{-1} \d (G \phi) ] + G^{-1} \d (G \, \delta \phi) = 0 \; .
\ee
Again, the game is to find $\d$ and $\delta$ such that this reduces
to an interesting equation for $G$ independent of $\phi$.
The following subsections present several examples.

\subsection{A unifying bicomplex framework for some integrable models} 
Let $M^0$ be the space of $2 \times 2$ matrices with entries in 
$C^\infty (\mathbb{R}^3)$ and $M = M^0 \otimes \Lambda_2$. In terms of 
coordinates $t,x,y$ on $\mathbb{R}^3$ we define linear maps 
$\d , \delta \, : \, M^0 \rightarrow M^1$ via
\be
  \d \phi = \phi_t \, \tau + \phi_x \, \xi \, , \quad
  \delta \phi = \phi_y \, \tau + {1 \over 2i} \, (I - \sigma_3)  \phi \, \xi
\ee
where $I$ is the $2 \times 2$ unit matrix and $\sigma_3 = \mbox{diag}(1,-1)$. 
The bicomplex conditions are then identically satisfied. 
The $\delta$-cohomology is not trivial. For example, elements of the form
$(c(t,x),0) \, \xi$ are $\delta$-closed but not $\delta$-exact.
\vskip.2cm

Now we dress $\d$ with some invertible $2 \times 2$ matrix $G$ as follows,
\be
    \D \phi = G^{-1} \d (G \phi) 
  = (\phi_t - V \, \phi) \, \tau + (\phi_x - U \, \phi) \, \xi 
\ee
where 
\be
    U = - G^{-1} G_x \, , \qquad
    V = - G^{-1} G_t   \; .
\ee 
In terms of $U$ and $V$, the bicomplex equation $\D^2 = 0$ reads
\be
     U_t - V_x + [ U , V ] = 0   \label{U-V-eq1}
\ee
which is an identity in the case under consideration.
The only nontrivial bicomplex equation is $\delta \D + \D \delta = 0$ which 
takes the form
\be
   U_y - {i \over 2} \, [ \sigma_3 , V ] = 0 \; .  
                \label{U-V-eq2}
\ee
\vskip.2cm
\noindent
{\em Example.} Let 
\be
   G = \exp \left( {i \over 2} \, (I-\sigma_3) \, t \right) \, 
       \exp \left( {i \over 2} \, \sigma_2 \, u \right) 
       \, , \qquad
   \sigma_2 = \left( \begin{array}{cc} 0 & -i \\
                                       i &  0 
                     \end{array} \right)
\ee 
where $u$ does not depend on $t$. Then
\be
   U = \left( \begin{array}{cc} 
                       0 & -u_x/2 \\
                   u_x/2 & 0  
              \end{array} \right) 
       \, , \quad
   V = \left( \begin{array}{cc} 
                - \sin^2(u/2) & \sin(u/2) \, \cos(u/2) \\
       \sin(u/2) \, \cos(u/2) & - \cos^2(u/2)  
              \end{array} \right) \, ,
\ee
and (\ref{U-V-eq2}) is equivalent to the sine-Gordon equation 
$u_{xy} = \sin u$.
          {  }   \hfill   \rule{5pt}{5pt}
\vskip.2cm

Now we decompose $V$ as follows,
\be
    V = i \, ( V^+ + V^- ) \, \sigma_3    \label{V-decomp}
\ee
where
\be
    \sigma_3 \, V^+ \, \sigma_3 = V^+ \, , \qquad
    \sigma_3 \, V^- \, \sigma_3 = - V^-  \; .
\ee
Then (\ref{U-V-eq2}) becomes
\be
     V^- = U_y  \; .                        \label{V-}
\ee
This implies $\sigma_3 \, U_y \, \sigma_3 = - U_y$ which restricts $U$ 
to the following form,
\be
    U = \left( \begin{array}{cc} 0 & q \\ r & 0 \end{array} \right)
                \label{U-AKNS}
\ee
with functions $q$ and $r$, up to addition of terms on the diagonal which 
do not depend on $y$. If the latter vanish, then 
\be
   \sigma_3 \, U \, \sigma_3 = - U    \; .
                             \label{U-sig3-antic}
\ee
Using (\ref{V-}) to eliminate $V^-$ from (\ref{U-V-eq1}), we
obtain the two equations
\be
    V^+ = \partial_x^{-1} (U^2)_y     \label{V+}
\ee 
and
\be
   i \, U_t = - [ U_{xy} - 2 \, U \, \partial_x^{-1} (U^2)_y ] 
               \, \sigma_3 \; .             \label{U-eq}
\ee
Here $\partial_x^{-1}$ means integration with respect to $x$. 
In conclusion, we note the following. Let $U$ be of the form 
(\ref{U-AKNS}). Defining $V$ via 
(\ref{V-decomp},\ref{V-},\ref{V+}), then the bicomplex 
equations are satisfied iff (\ref{U-eq}) holds.
(\ref{U-eq}) generalizes the nonlinear Schr\"odinger equation. Indeed, 
setting $y=x$ (which reduces the system to two dimensions) and choosing 
$q = \bar{\psi}$ and $r = \psi$ with a complex function $\psi$ with 
complex conjugate $\bar{\psi}$, the equation (\ref{U-eq}) 
becomes equivalent to $i \, \psi_t = - \psi_{xx} + 2 \, |\psi|^2 \, \psi$ 
(and its complex conjugate). 
Relations with the AKNS formalism are rather obvious 
(cf \cite{Das89}, for example).
\vskip.2cm

Suppose we have an invertible matrix, say $\tilde{\chi}$, the entries of 
which are functions of two coordinates, say $t$ and $x$. Then the identity
\be
  ( \tilde{\chi}_x \tilde{\chi}^{-1} )_t = ( \tilde{\chi}_t \tilde{\chi}^{-1} )_x
  + [ \tilde{\chi}_t \tilde{\chi}^{-1} , \tilde{\chi}_x \tilde{\chi}^{-1} ] 
\ee
implies that 
\be
    \mbox{tr} ( \tilde{\chi}_x \tilde{\chi}^{-1} )_t 
  = \mbox{tr} ( \tilde{\chi}_t \tilde{\chi}^{-1} )_x
\ee
which has the form of a conservation law. However, the nontrivial task is 
to ensure that the brackets on both sides contain only terms which 
are {\em local}\footnote{In the sense of not involving integrals of the field.} 
in  the field we are interested in (see also \cite{Wils81}). 
\vskip.2cm

Let us now turn to the associated linear system $\delta \chi = \lambda \, \D \chi$
for a matrix $\chi \in M^0$, i.e.,
\be
       \chi_y = \lambda \, (\chi_t - V \, \chi) \, , \qquad
   e_{-} \chi = i \, \lambda \, (\chi_x - U \, \chi)
\ee
where $e_{\pm} = (I \pm \sigma_3)/2$. The second equation implies 
\be
   e_{+} \chi_x = e_{+} U \chi  \label{e+chi_x}
\ee
and, using (\ref{U-sig3-antic}), 
\be
      e_{+} \chi_x 
    = U \, e_{-} \chi
    = i \lambda ( U \, \chi_x - U^2 \chi) \; .
\ee
On the other hand, differentiating the second equation of the linear
system with respect to $x$ leads to
\be
   e_{-} \chi_x = i \, \lambda \, (\chi_{xx} - U_x \, \chi - U \, \chi_x)
\ee
and combining the last two equations gives
\be
   \chi_x = i \, \lambda \, ( \chi_{xx} - (U_x + U^2) \, \chi )  \; .
            \label{chi-x-eq}
\ee
Assuming that $U$ is invertible (i.e., $q r \neq 0$), (\ref{e+chi_x}) 
can also be written as
\be
     e_{-} \chi = e_{-} U^{-1} \chi_x   \; .
\ee
The linear system implies $e_{-} \chi^{(0)} = 0$ which is solved by 
$\chi^{(0)} = e_{+}$. In particular, it follows that $\chi$ is 
not invertible (as a formal power series in $\lambda$). 
Let us consider instead 
\be
     \tilde{\chi} = \chi + e_{-} 
\ee
which is invertible.  
Using
\be
  \chi = e_{+} \chi + e_{-} \chi
       = e_{+} \tilde{\chi} + e_{-} U^{-1} \tilde{\chi}_x
\ee
(\ref{chi-x-eq}) becomes
\be
   \tilde{\chi}_x = i \, \lambda \, ( \tilde{\chi}_{xx} 
   - (e_{-} U + e_{+} U_x U^{-1} ) \, \tilde{\chi}_x 
   - (e_{-} U_x + e_{+} U^2) \, \tilde{\chi} )  \; .
\ee
Introducing $\theta$ via
\be
    \tilde{\chi}_x \tilde{\chi}^{-1} = \lambda \, \theta
\ee
the last equation takes the form
\be
    \theta = -i \, ( e_{-} U_x + e_{+} U^2 )
    + i \, \lambda \, ( \theta_x - (e_{-} U + e_{+} U_x U^{-1} ) \, \theta )
    + i \, \lambda^2 \, \theta^2     \; .    \label{theta-eq}
\ee
Inserting the power series expansion 
\be
    \theta = \sum_{k \geq 0} \lambda^k \, \theta^{(k)}
\ee
in (\ref{theta-eq}), we find 
\be
    \theta^{(0)} = - i \, (e_{-} U_x + e_{+} U^2) \, , \qquad 
    \theta^{(1)} = e_{+} U U_x + e_{-} (U_{xx} - U^3)
\ee 
and
\be
    \theta^{(k)} =  i \, ( \theta^{(k-1)}_x 
   - (e_{-} U + e_{+} U_x U^{-1} ) \, \theta^{(k-1)} ) 
   + i \sum_{j=0}^{k-2} \theta^{(j)} \, \theta^{(k-2-j)}   
\ee
for $k > 1$. According to the general argument given above, the quantities
\be
     w^{(k)} = \mbox{tr} \, \theta^{(k)}
\ee
are conserved in a two-dimensional sense (with respect to both coordinate 
pairs $t,x$ and $y,x$). A more explicit form of the conservation laws is given 
in the addendum below. We find
\be
    w^{(0)} = -i \, \mbox{tr} ( e_{+} U^2 ) = -i \, q \, r  \, , \quad
    w^{(1)} = \mbox{tr} (e_{+} U U_x) = q \, r_x
\ee
and
\be
   w^{(k)} =  i \, ( w^{(k-1)}_x 
   - {q_x \over q} \, w^{(k-1)} ) 
   + i \sum_{j=0}^{k-2} w^{(j)} \, w^{(k-2-j)}   \; .
\ee
With $y=x$ and $q = \bar{\psi}$, $r = \psi$, one recovers from the last 
equations the conserved densities of the nonlinear Schr\"odinger equation.  
Choosing $r=-q=u_x/2$ where $u=u(x,y)$, the $w^{(k)}$ reproduce the conserved 
densities of the sine-Gordon equation.    \\
{\em Addendum.} 
The first equation of the linear system can be rewritten as follows,
\be
    \tilde{\chi}_y \tilde{\chi}^{-1} 
 = \lambda \, (\tilde{\chi}_t \tilde{\chi}^{-1} 
   - V \, (e_{+} + \lambda \, e_{-} U^{-1} \theta )  \; .
\ee
With its help we get
\be
     \theta_y 
 &=& {1 \over \lambda} ( \tilde{\chi}_x \tilde{\chi}^{-1} )_y
  = {1 \over \lambda} (\tilde{\chi}_y \tilde{\chi}^{-1})_x
    +  [ \tilde{\chi}_y \tilde{\chi}^{-1} , \theta ]  \nonumber \\
 &=& ( \tilde{\chi}_t \tilde{\chi}^{-1} 
    - V \, (e_{+} + \lambda \, e_{-} U^{-1} \theta ) )_x
    + \lambda [ \tilde{\chi}_t \tilde{\chi}^{-1} 
    - V \, (e_{+} + \lambda \, e_{-} U^{-1} \theta ) , \theta ]  \; .
\ee
Similarly we obtain
\be
     \theta_t 
 &=& {1 \over \lambda} ( \tilde{\chi}_x \tilde{\chi}^{-1} )_t
  = {1 \over \lambda} (\tilde{\chi}_t \tilde{\chi}^{-1})_x
    +  [ \tilde{\chi}_t \tilde{\chi}^{-1} , \theta ]  \nonumber \\
 &=& ( \lambda^{-2} \tilde{\chi}_y \tilde{\chi}^{-1} 
     + \lambda^{-1} V \, (e_{+} + \lambda \, e_{-} U^{-1} \theta ) )_x
     + [ \lambda^{-1} \tilde{\chi}_y \tilde{\chi}^{-1} 
     + V \, (e_{+} + \lambda \, e_{-} U^{-1} \theta ) , \theta ]  \, .  \qquad
\ee
It follows that $w = \mbox{tr} \, \theta$ satisfies the conservation laws
\be
  w_y &=& \mbox{tr}( \tilde{\chi}_t \tilde{\chi}^{-1}
          - V \, ( e_{+} + \lambda \, e_{-} U^{-1} \theta ) )_x  \\
  w_t &=& \mbox{tr}( \lambda^{-2} \tilde{\chi}_y \tilde{\chi}^{-1} 
          + \lambda^{-1} V \, (e_{+} + \lambda \, e_{-} U^{-1} \theta ) )_x
    \; . 
\ee
 {  }   \hfill   \rule{5pt}{5pt}
\vskip.2cm

If we apply a gauge transformation with $g = G^{-1}$
to the bicomplex associated with the nonlinear Schr\"odinger equation (where $y=x$), 
we obtain
\be
     \D' \phi &=& \phi_t \, \tau + \phi_x \, \xi \\
 \delta' \phi &=& G \, \delta (G^{-1} \phi) 
               = (\phi_x - G_x G^{-1} \phi) \, \tau 
                  + {1 \over 2i} \, (I - S)  \phi \, \xi
\ee
where $S = G \sigma_3 G^{-1}$.
The bicomplex conditions now take the form
\be 
    S_x = [ G_x G^{-1} , S ] \, , \qquad
    S_t = 2 \, i \, (G_x G^{-1})_x   \; . 
\ee
The first equation leads to $G_x G^{-1} = - (1/2) S S_x$ using
$G^{-1} G_x \, \sigma_3 + \sigma_3 \, G^{-1} G_x = 0$ 
(cf (\ref{U-sig3-antic})), and the second takes the form
\be
    S_t = -i \, (S S_x)_x 
        = -i \, ( S S_x - {1 \over 2} (S S_x + S_x S) )_x 
        = - {i \over 2} \, [ S , S_{xx} ]
\ee
where we used $S^2 = I$. This is the Heisenberg magnet equation
\be
      \vec{S}_t = \vec{S} \times \vec{S}_{xx}
\ee
where we have set $S = \vec{S} \cdot \vec{\sigma}$. We have thus reproduced 
the equivalence of the nonlinear Schr\"odinger equation and the 
Heisenberg magnet (cf \cite{Fadd+Takh87}, for example).

\subsection{Further bicomplex formulations of integrable models} 
{\em Sine-Gordon equation again.} For $z \, : \, \mathbb{R}^2 \to \mathbb{C}$ 
we define
\be
     \d z &=& {1 \over 2} (\bar{z}-z) \, \tau + z_x \, \xi  \\
 \delta z &=& z_t \, \tau + {1 \over 2}(\bar{z}-z) \, \xi
\ee
where $\bar{z}$ denotes the complex conjugate of $z$. 
Then $(M=C^\infty(\mathbb{R}^2,\mathbb{C}) \otimes \Lambda_2, \d, \delta)$
is a (trivial) bicomplex. Deforming $\d$ to
\be
   \D z = e^{-i \varphi/2} \d (e^{i \varphi/2} z)
        = {1 \over2} (e^{-i \varphi} \bar{z}-z) \, \tau 
		  + (z_x + {i\over2} \varphi_x \, z) \, \xi  \, ,
\ee  
$\delta \D + \D \delta = 0$ turns out to be equivalent to the 
sine-Gordon equation $\varphi_{tx} = \sin \varphi$.
The first cohomology group of $\delta$ is not trivial. In particular, $\xi$ 
is $\delta$-closed, but not in $\delta(M^0)$.
\vskip.1cm

With $\chi^{(0)}=1$ the linear equation $\delta \chi = \lambda \, D \chi$  
consists of the two equations
\be
   \chi_t = {\lambda \over 2}(e^{-i \phi}\,\bar{\chi}-\chi)  \, , \qquad
   {1\over2} (\bar{\chi}-\chi) &=& \lambda \, (\chi_x + {i \over 2} \phi_x \, \chi)
      \; .
\ee
Writing $\chi = \alpha + i \, \beta$, the second equation of the linear system
becomes
\be
   \alpha_x - {1\over2} \, \phi_x \, \beta = 0 \, , \qquad
   \beta = - \lambda \, (\beta_x + {1 \over 2} \, \phi_x \, \alpha) \; .
\ee
Eliminating $\beta$ and setting $\alpha = e^{- \lambda \gamma}$ with 
a function $\gamma$ yields
\be
    \gamma_x 
  = {1\over4} \phi_x{}^2 - \lambda \, (\gamma_{xx}-{\phi_{xx} \over \phi_x} \gamma_x)
    + \lambda^2 \, \gamma_x{}^2  \; .
\ee
 From the first equation of the linear system we obtain
\be
   \alpha_t = {\lambda \over 2} (\alpha \cos \varphi-\beta \sin \varphi - \alpha)
       \; .
\ee
After some simple manipulations one arrives at the conservation law
\be
  (\gamma_x)_t = - (\lambda \, {\sin \varphi \over \varphi_x} \, \gamma_x
  + {1 \over 2} \cos \varphi)_x  \; .
\ee
Inserting the power series expansion for $\gamma$ with respect to $\lambda$, 
one obtains the conserved quantities
\be
   \gamma_x^{(0)} = {1 \over 4} \phi_x{}^2 \, , \quad
   \gamma_x^{(1)} = - {1 \over 8} (\phi_x{}^2)_x \, , \quad \ldots
\ee
of the sine-Gordon equation. Similar calculations can be performed in case of 
the following models.
\vskip.2cm
\noindent
{\em Discrete sine-Gordon equation.}
Let $M^0$ be the space of complex functions on an infinite plane square lattice.
We define linear maps $\d , \delta \, : \, M^0 \rightarrow M^1$ by
\be
   (\delta z)_S &=& (z_E-z_S) \, \tau + a \, (\bar{z}_W-z_S) \, \xi  \\
   (\d z)_S &=& a \, (\bar{z}_E-z_S) \, \tau + (z_W-z_S) \, \xi 
\ee
where an index $N,S,E,W$ means: take the value at the north, south, east 
and west point, respectively, of a common elementary square of the lattice 
(cf \cite{Fadd+Volk94} for this notation). 
It is easily verified that $(M=M^0 \otimes \Lambda_2,\d,\delta)$ is a bicomplex.
Now we deform $\d$ to
\be
   (\D z)_S = e^{-i \varphi_S/2} ( \d (e^{i \varphi/2} z) )_S
   = a \left(e^{-i(\varphi_S+\varphi_E)/2} \bar{z}_E-z_S \right) \, \tau
     + \left(e^{-i(\varphi_S-\varphi_W)/2} z_W - z_S \right) \, \xi 
\ee
with a real function $\varphi$. Then $\delta \D + \D \delta = 0$ is equivalent to
\be
   e^{i(\varphi_N-\varphi_E)/2}-e^{-i(\varphi_S-\varphi_W)/2}
   = a^2 \left(e^{i(\varphi_N+\varphi_W)/2}-e^{-i(\varphi_S+\varphi_E)/2} \right) \; .
\ee
Taking real and imaginary parts and using some trigonometry, one proves that this
is equivalent to
\be
  \sin \left[{1\over 4}(\varphi_N+\varphi_S-\varphi_E-\varphi_W)\right]=
  a^2 \sin \left[{1\over 4}(\varphi_N+\varphi_S+\varphi_E+\varphi_W) \right]
\ee
which is the discrete sine-Gordon equation \cite{Hiro77}.

\vskip.2cm
\noindent
{\em Toda field theory.} Let $M^0$ be the algebra of functions on 
$\mathbb{R}^2 \times \mathbb{Z}$ which are smooth in the first two arguments.
We write $f_k(t,x) = f(t,x,k)$ for $k \in \mathbb{Z}$.
Supplied with the linear maps defined by
\be
  \delta f_k = (\pa_t f_k) \, \tau + (f_{k+1}-f_k) \, \xi \, , \quad
  \d f_k = -(f_k-f_{k-1}) \, \tau + (\pa_x f_k) \, \xi
\ee
$M = M^0 \otimes \Lambda_2$ becomes a bicomplex. The bicomplex conditions 
(\ref{bicomplex_cond}) are identically satisfied. Now we dress $\d$ as follows,
\be
   \D f_k = e^{-q_k} \d (e^{q_k} f_k) 
          = (e^{q_{k-1}-q_k} f_{k-1}-f_k) \, \tau
		    + (\pa_x f_k + (\pa_x q_k) f_k) \, \xi 
\ee
where $q \in M^0$.
Then $\delta \D + \D \delta = 0$ is equivalent to the Toda field equation
\be
   \pa_t \pa_x q_k = e^{q_k-q_{k+1}} - e^{q_{k-1}-q_k}  \; .
\ee               
See also \cite{DMH00a,DMH99c,DMH00b} for some related work.
\vskip.2cm
\noindent
{\em A generalization of Hirota's difference equation.}
Hirota's difference equation is a discretization of the Toda field theory 
\cite{Hiro81}. Let $M^0$ be the algebra of functions of $n$ discrete variables 
$x^1,\ldots,x^n$ and $M = M^0 \otimes \Lambda_n$. $M$ becomes a bicomplex with
$\d$ and $\delta$ determined by 
\be
   \d f = \sum_i a_i \, (R S_i f -f) \, \xi^i  \, , \qquad
   \delta f = \sum_i b_i \, (S_i f-f) \, \xi^i 
\ee
where $(S_i f)(x^1,\ldots,x^n) = f(x^1,\ldots,x^{i-1},x^i+1,x^{i+1},\ldots,x^n)$. 
$R$ is an automorphism of the algebra of functions commuting with $S_i$, and 
$a_i, b_i$ are constants. The bicomplex equations are then identically satisfied. 
Now we deform $\d$ to
\be
  \D f = e^{-q} \, \d (e^q f) 
       = \sum_i a_i \, \left( e^{R S_i q -q} \, R S_i f - f \right) \, \xi^i 
\ee
with a function $q$ of the discrete variables. Then $\delta \D + \D \delta = 0$ 
yields
\be
   a_i b_j \, \left( e^{R S_i S_j q-S_j q} - e^{R S_i q-q} \right)
  = a_j b_i \, \left( e^{R S_i S_j q-S_i q} - e^{R S_j q-q} \right)  \; .
\ee

\subsection{Remarks on self-dual Yang-Mills equations and reductions}
Let $\A$ be an associative (and not necessarily commutative) algebra over 
$\mathbb{R}$ or $\mathbb{C}$. Furthermore, let $M= \O \otimes_\A \A^m$ where 
$(\O,\d,\delta)$ is a bidifferential calculus over $\A$ (cf \cite{DMH00a}), 
so that $\d$ and $\delta$ satisfy the (graded) Leibniz rule. 
If we apply dressings of the second kind to {\em both} generalized exterior 
derivatives, we obtain a bicomplex $(M,\D,\Dc)$ where
\be
   \D \phi = \d \phi + A \, \phi \, , \qquad
   \Dc \phi = \delta \phi + B \, \phi         \label{tc}
\ee
if the bicomplex conditions
\be
    \d A + A^2 = 0 \, , \quad
    \delta B + B^2 = 0  \, , \quad
    \d B + \delta A + A B + B A = 0 
\ee
are satisfied. The first two equations are solved, of course, with 
$A = G^{-1} \d G$ and an analogous expression for $B$. 
Applying suitable gauge transformations, we 
obtain equivalent bicomplexes $(M, \D', \delta)$ and $(M, \d, \Dc')$, in
particular.
\vskip.2cm

Let us now specialize $\A$ to the commutative algebra of smooth functions 
of $2n$ variables $x^i, y^i$, $i=1,\ldots,n$, and set
\be
    \d f =\sum_i {\pa f \over \pa x^i} \, \xi^i \, , \qquad
    \delta f = \sum_i {\pa f \over \pa y^i} \, \xi^i
\ee
where $\xi^i,\, i=1,\ldots,n$, is a basis of $\Lambda_n$. Since $B$ can be 
transformed to zero by a gauge transformation, it is sufficient to consider 
$(M,\D,\delta)$ which is a bicomplex iff 
\be
    \d A + A^2 = 0 \, ,  \qquad 
	\delta A = 0  \; .
\ee
 For $n=2$ this is gauge equivalent to the self-dual Yang-Mills equations 
\cite{Yang77,DMH00a}. 
Higher-dimensional generalizations of the above kind with $n>2$ have been
considered in \cite{Ward84,DMH00a}. 
Many examples of integrable models can be obtained from (\ref{tc}) via a 
{\em reduction}. This means that one considers cases where the fields depend 
only on particular combinations of the variables $x^i,\,y^j$ and the 
connection 1-forms have special forms (cf \cite{sdYM_red}). Since reductions do
not commute with gauge transformations, it is necessary, however, to consider 
more generally the bicomplex $(M,\D,\Dc)$ where also $B$ is switched on.

\vskip.2cm
\noindent
{\em Example.} Let the functions depend on the $x^i$, $i=1, \ldots,n$, only. 
With $A=0$ we obtain
\be
   \d \phi = \pa_i \phi \, \d x^i \, , \qquad
  \Dc \phi = B_i \phi \, \d x^i
\ee
and the bicomplex conditions read
\be
   \pa_i B_j -\pa_j B_i = 0  \, , \qquad
   B_i B_j -B_j B_i = 0 \; .
\ee
 For $m=n$ and assuming that there is a constant metric tensor
$\eta_{ij}$ such that for $B_k = (B_k{}^i{}_j)$ the tensor 
$B_{ijk} = \eta_{jl} B_i{}^l{}_k$ is totally symmetric, the above 
equations are equivalent to the WDVV equations \cite{Dubr92}.  
  {  }   \hfill   \rule{5pt}{5pt}

\section{Conclusions}
\setcounter{equation}{0}
By application of gauge transformations and dressings we have constructed 
bicomplex zero curvature formulations for several integrable models.
The associated linear system then arises in a form with a linear dependence 
on the (spectral) parameter $\lambda$ and the existence of an infinite set 
of conserved densities follows from the general recursive construction of 
$\delta$-closed elements in the bicomplex associated with the respective model 
\cite{DMH00a}. A relation between our bicomplex formulation and (finite-dimensional) 
bi-Hamiltonian systems has recently been revealed in \cite{CST00}.
\vskip.2cm

The general iterative construction applies to a much wider range of (weak) 
bicomplexes than those related to classical soliton equations and integrable 
models. In particular, generalizations of classical integrable models to
corresponding models on noncommutative spaces are obtained in this framework 
by replacing the ordinary product of functions by the Moyal 
$\ast$-product \cite{DMH00b} (see also \cite{Seib+Witt99} and the references 
cited there).

\end{document}